# Ultrafast quantum light uncertainty dynamics in real time

Mohamed Sennary[1], Javier Rivera-Dean[2], Mohamed ElKabbash[3], Vladimir Pervak[4], Maciej Lewenstein[2,5], Mohammed Th. Hassan*[1,3].

[1] Department of Physics, University of Arizona, Tucson, AZ 85721, USA.

[2] ICFO–Institut de Ciencies Fotoniques, The Barcelona Institute of Science and Technology, Castelldefels (Barcelona) 08860, Spain.

[3] James C. Wyant College of Optical Sciences, University of Arizona, Tucson, Arizona 85721, USA

[4] Ludwig-Maximilians-Universität München, Am Coulombwall 1, 85748, Garching, Germany.

[5] ICREA, Pg. Lluis Companys, 23, 08010 Barcelona, Spain.

*Corresponding author. Email: mohammedhassan@arizona.edu.

**Advancements in quantum optics and squeezed light generation have transformed various domains of quantum science and technology[1-5]. However, real-time quantum dynamics remain an underexplored frontier. Here, we extend quantum optics into the ultrafast regime, providing direct experimental evidence that quantum uncertainty is not a static constraint but evolves dynamically with the system's state and interactions. Using ultrafast squeezed light generated via a four-wave mixing nonlinear process[6,7], we observe the temporal dynamics of amplitude uncertainty, demonstrating that quantum uncertainty is a controllable and tunable physical quantity. This offers new insights into fundamental quantum mechanics in real time. Additionally, we demonstrate control over the quantum state of light by switching between amplitude and phase squeezing. Our ability to generate and manipulate ultrafast squeezed light waveforms with attosecond resolution unlocks exciting possibilities for quantum technologies, including petahertz-scale secure quantum communication, quantum computing, and ultrafast spectroscopy. We also introduce an ultrafast quantum encryption protocol leveraging squeezed light for secure digital communication at unprecedented speeds. This work paves the way for exploring quantum uncertainty dynamics and establishes the foundation for the emerging field of ultrafast quantum science.**

The uncertainty principle is a fundamental concept in quantum mechanics typically understood as a static constraint. Experimental investigation of the uncertainty dynamic has remained elusive, primarily due to the lack of tools capable of probing quantum phenomena on ultrafast timescales. Advancements in ultrafast optics, recognized by the 2018 and 2023 Nobel Prizes in Physics, have enabled the generation of laser pulses with attosecond durations—based on a high-harmonic-generation (HHG) process—providing unprecedented temporal resolution for studying light-matter interactions. These advancements have revolutionized our understanding of electron dynamics in atoms, molecules, and solids.[8-12] However, these studies have primarily focused on classical light sources, leaving the ultrafast dynamics of quantum light largely unexplored.

Squeezed light,[7] a non-classical state with reduced noise in one quadrature at the expense of increased noise in the conjugate quadrature, has been used in enhancing the sensitivity of gravitational wave detectors and enabling secure quantum communication.[1-6,13] Recent experimental and theoretical studies have explored the impact and generation of squeezed light on the HHG process, revealing that the underlying physics of HHG with squeezed light deviates significantly from classical descriptions[14-23]. Theoretical models predict shifts in both the electronic trajectories driven by the field and the emitted photon energy compared to those observed under coherent laser fields[24,25]. Experimental demonstrations of HHG[23,26] in solid-state materials using squeezed light have also shown promising results, yielding higher photon fluxes.[14,27] Additionally, the control of electron and XUV photon statistics through different quantum light modes has been reported, suggesting the potential for tailored quantum control in ultrafast processes.[23,28]

In this work, we bridge the gap between ultrafast science and quantum optics by generating and characterizing ultrafast squeezed light pulses in a light field synthesizer[29]. This technique allows us to control the temporal and spectral properties of the squeezed light, enabling us to probe its real-time dynamics with attosecond resolution. Our experimental results reveal that the uncertainty in amplitude-squeezed light is dynamic, varying with the system's state and its interactions. This ability to manipulate and observe the change of uncertainty in real time opens new avenues for exploring the fundamental nature of quantum mechanics and its implications for quantum control-based technologies. Furthermore, our ability to generate and control ultrafast squeezed light opens exciting new possibilities for quantum technologies. Accordingly, we demonstrate the potential of this technology for petahertz-scale secure quantum communication, exceeding the capabilities of classical communication systems.

This new perspective in understanding could potentially lead to a revised interpretation of quantum mechanics, where uncertainty is not merely a limitation but a dynamic resource that can be harnessed and controlled. Also, the implications of our findings extend beyond fundamental quantum mechanics to the technology by using ultrafast squeezed light for advancing quantum computing and ultrafast spectroscopy, offering new tools to probe and control the quantum world on unprecedented timescales.

**Ultrafast squeezed light**

Harnessing the quantum properties of light for ultrafast applications requires the generation of squeezed light pulses with precise temporal and spectral control. In this study, we achieve this by combining a light field synthesizer (LFS)[29-33]. with a four-wave mixing (FWM) process in a nonlinear crystal. The LFS allows us to manipulate the spectral phase and amplitude of a broadband laser pulse, enabling the synthesis of complex waveforms with attosecond precision.

By driving the FWM process with these tailored pulses, we generate squeezed light states that inherit the ultrafast temporal structure of the LFS output. Figure 1a illustrates the experimental setup. A broadband supercontinuum, generated by propagating an ultrafast laser pulse through a hollow-core fiber filled with neon gas, is split into three spectral channels: near-infrared (NIR, 1000-690 nm), visible (VIS, 715-500), and ultraviolet (UV, 515-350 nm). Each channel is compressed using chirped mirrors to approach its Fourier limit, resulting in pulse durations of approximately 8-10 fs (Fig. S1). These pulses are then recombined using dichroic beam splitters and coherently superimposed at the output of the LFS, forming a broadband few-cycle laser pulse. The relative delays and intensities of the constituent pulses can be precisely controlled, allowing for the synthesis of arbitrary waveforms. The LFS output is then split into two beams. One beam is directed to a spectrometer for spectral analysis (Fig. 1b, c), while the other beam is focused by two D-shaped focusing mirrors (one of them is controlled by a highly precise piezoelectric delay stage) onto a 100 μm-thick $SiO_2$ target, where FWM generates squeezed light in the amplitude quadrature. The conversion efficiency is estimated to be on the order of 0.1%, depending on the nonlinear material, thickness, and input beam intensity. The squeezed light is isolated through a spatial filter and recorded by a second spectrometer (Fig. 1d, e). To confirm the presence of amplitude squeezing, we employ a method based on spectral interference between the temporally delayed constituent pulses (NIR-VIS and VIS-UV). By analyzing the interference fringes within the spectral overlap region (insets in Fig. 1c and e), we extract the phase uncertainty ($\Delta\Phi$). The intensity uncertainty (ΔI) is obtained by measuring the variance in the overall pulse intensity. Amplitude squeezing is confirmed when the intensity fluctuations of the squeezed light ($\Delta I_s$) are smaller than those of the coherent light ($\Delta I_c$), while the phase jittering of the squeezed light ($\Delta\Phi_s$) exceeds that of the coherent light ($\Delta\Phi_c$).

Our measurements, summarized in Fig. 2, demonstrate that the generated light indeed exhibits amplitude squeezing. The phase uncertainty is higher in the squeezed light (Fig. 2aII and cII) compared to the coherent light (Fig. 2aI and cI), while the amplitude of the squeezed light (Fig. 2bII and dII) is more stable than that of the coherent light. These results confirm that the FWM process generates amplitude-squeezed light at the expense of increased phase uncertainty. To assess amplitude stability for individual pulse components, we spectrally filtered the constituent pulses and retrieved their intensity stability. The results shown in Fig. S2 indicate that the amplitude stability difference between the squeezed and classic light of the different channels is roughly comparable.

We also perform a shot noise measurement for both classic and squeezed light by measuring the standard deviation of I ($\sigma I$) as a function of the beam power and show it in Extended Data Fig. 1. This result is another proof of the obtained amplitude squeezing of ultrafast light.

Next, we performed theoretical modeling to investigate the nature of the measured phase and amplitude uncertainties are compare it with those of a squeezed state. Then, we calculated the Wigner function as a visual representation of these generated states. The Wigner function is a quasi-probability distribution that provides a visual representation of the quantum state in phase space. Our analysis reveals that the squeezed states correspond to a Gaussian distribution that has been squeezed along one optical quadrature and expanded along the conjugate quadrature (Fig. 3, see SI).

From a theoretical point of view, the four-wave mixing process under co-linear phase matching conditions and within the parametric approximation [7] produces squeezed coherent states of the form $\otimes_{i=1}^{N} \hat{S}(r_i)|\alpha_i\rangle$. In this expression, $N$ denotes the number of modes, $S(r) = exp[ra^2 - r^* a^{\dagger 2}]$ is the squeezing operator, and $|\alpha\rangle$ represents a coherent state of light. To determine the compatibility of the experimental results with these theoretical squeezed states, we compared the variances in phase and intensity obtained experimentally, $\Delta\Phi_{exp}^2$ and $\Delta I_{exp}^2$, with the theoretical predictions, $\Delta\Phi_{th}^2(\boldsymbol{x})$ and $\Delta I_{th}^2(\boldsymbol{x})$, (see Ref. [34] and the supplementary information (SI) for detailed calculations), the latter parametrized by $\boldsymbol{x} = \{(r_i, \alpha_i)\}_{i=1}^{N}$. Specifically, we defined the following function

$$C(\boldsymbol{x}) = A[\Delta I_{th}^2(\boldsymbol{x}) - \Delta I_{exp}^2]^2 + B[\Delta\Phi_{th}^2(\boldsymbol{x}) - \Delta\Phi_{exp}^2]^2, \quad (1)$$

which serves as a measure of the distance between the experimental observations and the theoretical expectations, where a value of $C(\boldsymbol{x}) = 0$ would represent a perfect match between the two. Consequently, the comparison involves finding an optimal set of parameters $\boldsymbol{x}^*$ that minimizes Eq. (1) as much as possible. To achieve this, the parameters $A$ and $B$ have to be chosen carefully to ensure simultaneous minimization of both terms within the function. In practice, the optimization was performed across a range of values for $A$ and $B$, and the best result was retained (see SI for more details). Furthermore, given that the experimental data suggests the presence of amplitude-squeezing, we restricted our search-space to real values of $r_i$ and $\alpha_i$ when solving this optimization problem, as to reduce the number of variables over which to optimize.

The results from this optimization are shown in Fig. 3a and Fig. 3b as a function of the number of modes $N$, where Fig. 3a represents the IR-Vis dataset and Fig. 3b the UV-VIS dataset. The dashed curves show the absolute difference between the experimental and theoretical variances after optimization, while the solid horizontal lines represent the experimental variance values. In all cases, we observe that the dashed curves lie below their respective solid horizontal lines—indicated in matching colors—demonstrating a good agreement between theory and experiment, as the absolute error remains around one and two orders of magnitude below the experimental values. Notably, this trend persists across all values of N considered in our study, suggesting that the experimental data is compatible with the presence of squeezed states along multiple modes. However, for both the IR-Vis and UV-VIS datasets, an increase in the number of modes generally corresponds to reduced accuracy in the absolute error, particularly pronounced in the intensity variance. Overall, we observe that the best results are achieved in the N=1 case, yielding a total estimated squeezing of 13.03 dB for the IR-Vis dataset and 8.81 dB for the UV-VIS dataset. The difference in the squeezing level may be attributed to the alignment optimization in two measurements to achieve the optimized FWM output signal. The Wigner function[35] of the resulting squeezed states are shown in Fig. 3c and Fig. 3d, respectively, where $\chi$ and $p$ represent the optical quadratures. The functions have been rotated by 45º relative to the $p$-axis for representational purposes. They correspond to a Gaussian distribution that has been squeezed along one optical quadrature and expanded along the conjugate quadrature.

**The squeezed light amplitude uncertainty dynamics in real time**

Having established the generation of ultrafast squeezed light pulse (a few femtoseconds synthesized squeezed light waveform is shown in Extended Data Fig. 2), we now turn our attention to the central question of this study: How does amplitude uncertainty is changing in real-time? To address this, we directly measure $\sigma I$, which presents the amplitude uncertainty of the squeezed light as a function of time. We exploit the temporal control between the three input beams in the FWM process to introduce a variable delay (τ) between one photon and the other two photons using the D-shape mirrors (see the setup in Fig.1a). By measuring $\sigma I$ for different values of τ, we can track the evolution of uncertainty as the system interacts and evolves in time. The measurements at different delay τ times are shown in Fig. 4a. Our results, presented in Fig. 4b (black circles connected by the black dashed line, and the red dashed line as a guide to the eye), show that the $\sigma I$ has higher values at the positive and negative delays in time. The minimum $\sigma I$ is achieved when all three input photons are overlapping time. These results demonstrate that $\sigma I$ is not static but varies in real-time, influenced by the system's state and its interactions.

This dynamical uncertainty of the amplitude can be explained as follows: the amplitude squeezing is achieved by utilizing the quantum correlations arising from the nonlinear four-wave mixing interaction among three input photons. In our case, the incident angle of these beams is kept very small (<5°). So, the good phase matching, which can be optimized by tilting the nonlinear medium, occurs and ensures constructive interference when the three photons arrive simultaneously in time. Consequently, the nonlinear signal generation window is temporally confined to the photons' overlap window ($\tau = -1.5\ to\ 1.5\ fs$, from Fig. 4b), reducing the uncertainty in the number and amplitude of generated photons. Conversely, if one of the input photons arrives with a certain delay (τ) relative to the other two, the nonlinear signal generation window widens, increasing randomness and, in turn, intensity uncertainty.

To further confirm this interpretation, we repeat the measurements with a squeezed phase light, obtained by altering the tilting angle of the nonlinear medium. The measured phase squeezing is shown in Extended Data Fig. 3a&b. In this case, the behavior of $\sigma I$ as a function of time, shown in Extended Data Fig. 3c (black circles connected by black dashed line, blue dashed line is a guide to the eye) is opposite to the case of amplitude squeezing (Fig. 4b), confirming that the dynamic behavior of uncertainty.

**Ultrafast quantum-encrypted digital communication**

The ability to generate and control ultrafast squeezed light pulses also unlocks exciting new possibilities for quantum technologies. One particularly promising application is ultra-secured quantum communication.

Traditional continuous-variable quantum key distribution (CV-QKD) relies on encoding information into the quadratures of quantum states, such as amplitude (ΔI) and phase ($\Delta\phi$), using a squeezed or coherent light source. The receiver measures one of these quadratures using a local oscillator (LO) to project the quantum state onto the desired quadrature, chosen at random. The security of CV-QKD arises from the inability of an eavesdropper (Eve) to intercept the quantum state without introducing detectable noise, owing to the Heisenberg uncertainty principle.

In our petahertz communication scheme, we leverage the ultrafast nature of the squeezed light pulses to encode quantum-encrypted digital data at unprecedented speeds. As illustrated in Fig. 5, Alice synthesizes and encodes digital data onto amplitude-squeezed light waveforms using the LFS, as we demonstrated earlier[29,30,36,37]. The measured exemplary waveforms shown in Fig. 5a I-III are the average of three scans, and despite the phase uncertainty of these waveforms, they maintain their shapes. Alice sets a predefined intensity threshold (30%) for the modulation, encrypting digital data within the squeezed waveforms. She then sends the encoded squeezed light beam to Bob (Fig. 5b), sharing the squeezing degree and the threshold information.

As in conventional CV-QKD, any eavesdropping attempt by Eve (Fig. 5c) to intercept and decode the squeezed light waveform would alter its squeezing degree, thereby revealing the intrusion to both Alice and Bob. Bob can check the signal's squeezing degree by measuring its intensity stability between different pulses (as we did in our experiment and shown in Fig. 2) to confirm the communication's security.

To further enhance security, we propose using one of the LFS pulses—specifically, the NIR pulse—as a quantum-correlated local oscillator [38]. This eliminates the need for an external classical LO and simplifies synchronization. At the receiver station, after sampling the waveform and decoding the digital data, Bob introduces a controllable relative time delay between the NIR reference pulse and the other pulses. This delay adjusts the interference conditions at specific wavelengths, allowing Bob to effectively select quadrature projections by inducing constructive or destructive interference. By varying the delay, Bob dynamically selects which quadrature component (amplitude or phase) to measure at a given wavelength. He then measures the noise at specific wavelengths within the spectral overlap region, where spectral interference exists between the different channels. Depending on the interference condition at that wavelength, Bob measures a specific quadrature and obtains the squeezing level of the selected quadrature projection. This protocol enhances security by making it difficult for an eavesdropper to predict Bob's quadrature choices or replicate the quantum-correlated states without introducing detectable noise. This approach avoids challenges associated with classical LOs, enables dynamic quadrature selection, and leverages quantum correlations to strengthen the security and efficiency of CV-QKD. Remarkably, our protocol introduces additional layers of security, in analogy to the typical CV-QKD protocol, protecting the communication from potential tampering. Specifically, our approach provides multiple safeguards against Eve's interference: (i) Eve would need the decoding key to correctly interpret the transferred information, (ii) Eve cannot accurately decode the data without knowing the predefined intensity threshold (30%), and (iii) even if Eve knows the threshold, her measurement of the squeezed light waveform using a beamsplitter will inevitably disturb the squeezing, introducing errors in the decoded data due to the altered squeezing degree. This disturbance increases the tolerance error of the predefined threshold and amplifies the likelihood of faulty decoding, as shown in Fig. 5 cI-III. Consequently, our approach not only secures the communication channel but also protects the transferred data from unauthorized retrieval.

This ultrafast quantum communication protocol operates at data transfer rates in the petahertz/second range, offering the potential for ultra-secure, high-speed communication networks. Although our petahertz quantum communication protocol offers a higher level of security due to the intrinsic sensitivity of squeezed light to any eavesdropping attempt, there are practical challenges, such as maintaining low-loss transmission of squeezed pulses over long distances. Scaling up the LFS and the input coherent laser pulse power to generate a high-power squeezing synthesized pulse would overcome the transmission loss problem and support robust ultrafast quantum communication. Furthermore, achieving quantum communication at petahertz frequency speeds is currently limited only by the laser repetition rate and the delay stage speed in the synthesizer. However, with ongoing advancements in ultrafast optoelectronics[39] and laser technologies, this digital data speed is potentially on the horizon.

## Discussion

This work introduces a groundbreaking approach to ultrafast quantum optics by demonstrating the generation of broadband squeezed light pulses with attosecond precision. By leveraging these ultrafast quantum light waveforms, we have shown how they can be used for secure quantum communication, marking a significant step toward the realization of high-speed, encrypted communication networks. Our results not only open new avenues for exploring quantum light-matter interactions in real-time but also lay the foundation for future advancements in ultrafast quantum optoelectronics and quantum computing.

As quantum communication systems evolve, the ability to operate at petahertz-scale data transmission speeds will be critical, and our findings contribute to the development of technologies that can meet these challenges while ensuring security and robustness. The potential of ultrafast squeezed light in secure communication represents a promising frontier for both fundamental research and practical applications in the quantum technology landscape. Moreover, the dynamic nature of quantum uncertainty observed in our experiments suggesting a more fluid picture of quantum reality. This has profound implications for understanding the fundamental nature of time and measurement in the quantum world. The ability to generate and control ultrafast squeezed light also opens exciting new possibilities for ultrafast spectroscopy and quantum computing. In ultrafast spectroscopy, this technology could be used to probe the dynamics of chemical reactions or biological processes on the attosecond timescale, providing unprecedented insights into the fundamental building blocks of matter. In quantum computing, ultrafast squeezed light could be used to manipulate qubits or perform quantum logic operations at unprecedented speeds, paving the way for the development of faster and more powerful quantum computers.

In conclusion, our work demonstrates the dynamic nature of quantum uncertainty, challenging the traditional static interpretation of the Heisenberg uncertainty principle. This discovery has far-reaching implications for our understanding of quantum mechanics and its applications across various scientific disciplines. This work paves the way to study the uncertainty dynamics as evolve in time in electronic wavepacket. Moreover, the ability to generate and control ultrafast squeezed light opens new possibilities for exploring the fundamental nature of quantum mechanics, developing novel quantum technologies, and advancing our understanding of the universe at its most fundamental level.

**Methods**

In our experimental setup, few-cycle laser pulses centered at 750 nm (passively stabilized carrier-envelope phase) are generated by an Optical Parametric Chirped-Pulse Amplification (OPCPA)-based laser system with a repetition rate of 20 kHz. These pulses propagate nonlinearly through a hollow-core fiber filled with neon gas (at 3.5 bar), generating a supercontinuum that spans from the ultraviolet to the near-infrared spectral range. The broadband spectrum is directed into a light field synthesizer (LFS), as shown in Fig. 1a, where it is split into three spectral channels using dichroic beam splitters: Ultraviolet (UV, 515-350 nm), Visible (VIS, 715-500 nm), and Near-Infrared (NIR, 1000-690 nm). Each of these channels is compressed using six pairs of chirped mirrors to approach their Fourier limits. The temporal profiles of the pulses in each channel are shown in Fig. S1. The full width at half maximum (FWHM) pulse durations are 10, 9, and 8.5 fs for the UV, VIS, and NIR pulses, respectively. The pulses are recombined using similar dichroic beam splitters and coherently superimposed at the output of the synthesizer, forming a two-cycle laser pulse. This pulse waveform is finely controlled by adjusting the relative delay between the three LFS pulses with a high-resolution piezoelectric linear stage in the UV and NIR beam paths. Neutral density filters are used to control the relative intensities of the constituent pulses. The total power of the output beam from the LFS is one watt.

The beam is first split using a beamsplitter, with the reflected beam (8% of the total power) representing the classic coherent synthesized pulse. This reflected beam is focused into spectrometer #1 (Ocean Optics HR 4000) with a 5-cm focal length lens. We measure 2800 spectra to assess the average intensity stability. First, we integrated each spectrum to get the total intensity ($I_n$), where n is the number of iterations. Then, $\Delta I$ was calculated by normalizing each spectrum to the average intensity of all the spectra and then calculating the percentage stability. The intensity standard deviation ($\sigma I$) percentage is calculated from the average and standard deviation of the n intensities for n iterations. On the other hand, the interference fringes between the spectral channels (inset of Fig. 1c and e) are analyzed, and their corresponding Fourier transforms are computed to extract phase jittering ($\Delta\Phi$) between the individual pulses.

The transmitted beam, after passing through the beamsplitter, is directed through a 3-hole mask. The three emerging beams, which are identical in power (162.5 mW), are focused using two D-shape focusing mirrors onto a fused silica sample (beam diameter ~50 μm). One of these D-shape mirrors is mounted on the piezo stage to control the relative delay between one of the input beams with respect to the other two beams. Then, a four-wave mixing process occurs, and a non-linear light signal is generated, which is amplitude-squeezed (153 $\mu$W). A 1-hole mask filters out this squeezed light, which is then focused into spectrometer #2 (Ocean Optics HR 4000CG) for characterization of its intensity, phase stability, and jitter. The LFS constituent pulses are delayed to generate better interference fringes at the spectral boundary of each pulse (see in set Fig. 1 c and e). Dark noise from the spectrometers (Fig. S3) was negligible, being two orders of magnitude lower than the observed amplitude fluctuations.

For the time-resolved intensity uncertainty measurements, we acquired 1000 spectra at each time delay (τ) between one of the input beams with respect to the other two beams, which interact with the $SiO_2$ to generate the nonlinear signal. We obtained $\sigma$I at each time delay τ and plotted it in Fig. 4.

Next, the relative delay between the LFS channels is also adjusted to synthesize the squeezed light waveform. The waveform is sampled using an all-optical light field sampling technique, reported previously [29,30,36,37]. This method involves focusing the squeezed light onto the $SiO_2$ sample (200 μm) alongside a probe beam. By measuring the modulation of the probe beam's transmission induced by the squeezed light, we extract the waveform by analyzing the recorded spectrum intensity as a function of the delay between the squeezed light and the probe pulse.

**Acknowledgments**:

This project is funded by the Gordon and Betty Moore Foundation Grant (GBMF 11476) to M. Hassan. This material is also based upon work partially supported by the Air Force Office of Scientific Research under award number FA9550-22-1-0494.

J.R.D. and M.L. acknowledge support from: European Research Council AdG NOQIA; MCIN/AEI (PGC2018 0910.13039/501100011033, CEX2019-000910 S/10.13039/501100011033, Plan National STAMEENA PID2022-139099NB), project funded by MCIN/AEI/10.13039/501100011033 and by the "European Union Next Generation EU/PRTR" (PRTR-C17.I1), FPI); QUANTERA DYNAMITE PCI2022-132919, Fundació Cellex; Fundació Mir-Puig.Fundació Cellex; Fundació Mir-Puig.

**Author contributions:**

M.S. conducted the experiments and analyzed the data. J.R.D and M.L. carried out the simulations and calculations. M.E. and M. Th. H. developed the ultrafast CV-QKD approach. M.E. suggesting the shot noise experiment. V.P. designed and measured the optics used in the setup. M. Th. H. conceived, supervised, and directed the study. All authors discussed the results and their interpretations and wrote the manuscript.

**Competing interests:**

Authors declare that they have no competing interests.

**Figures and Figure legends**

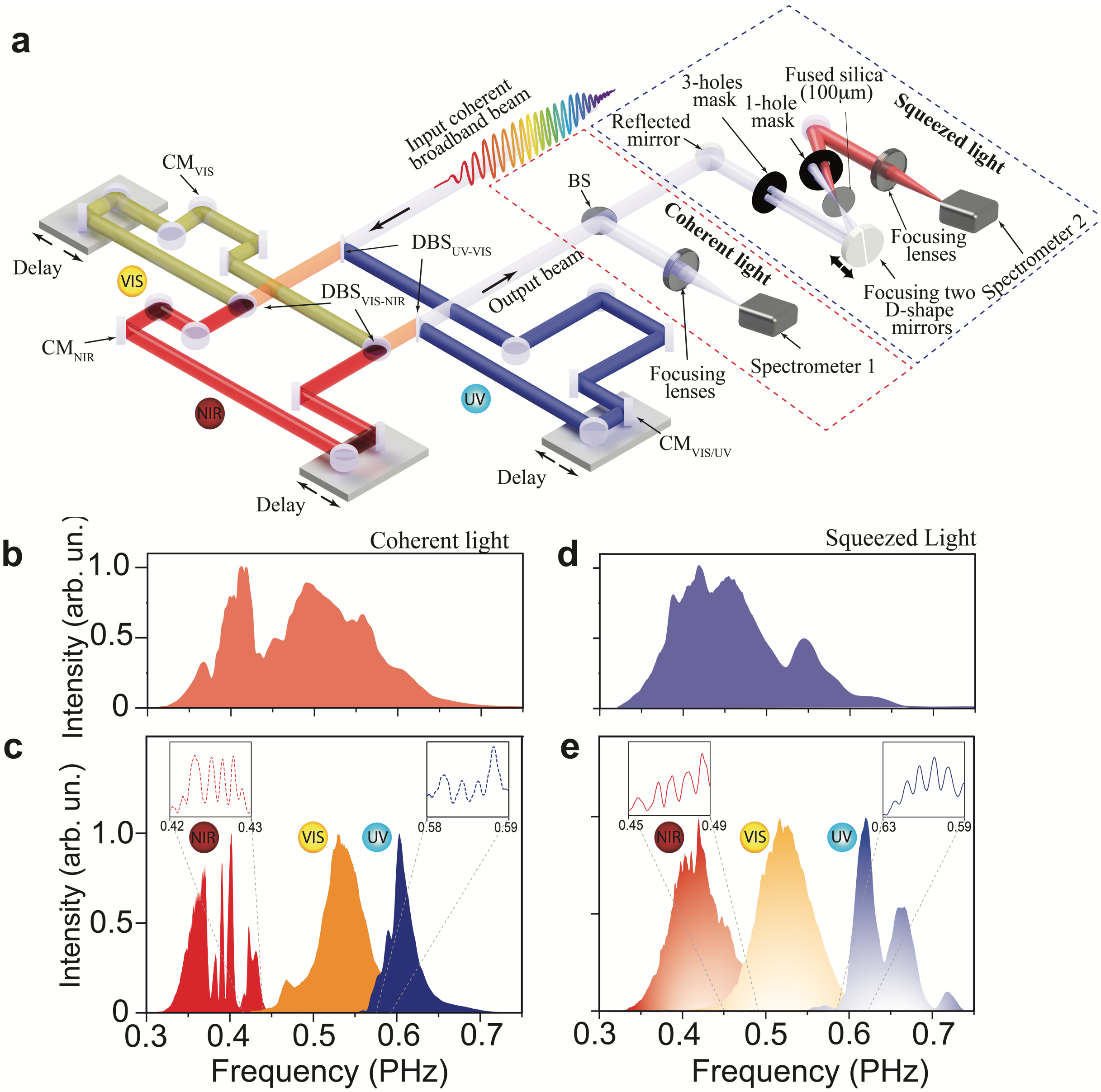

**Fig. 1. Generation of ultrafast squeezed light pulse synthesized with attosecond resolution. a**, Schematic of the light field synthesizer (LFS) setup, consisting of three spectral channels (pulses) used to generate synthesized waveform pulses. The output pulse is split into two beams: one is a classical coherent light pulse, serving as a reference, while the second beam undergoes a four-wave mixing process in a $SiO_2$ sample to generate a squeezed light pulse. The phase and intensity quadrature uncertainties of both beams are measured using spectrometers 1 and 2. **b&c**, Spectra of the broadband classical light pulse **(b)** and the spectra of its constituent LFS channels **(c)**. **d**, Spectrum of the generated squeezed light pulse. **e**, Spectra of the squeezed light generated by the three LFS pulses. Insets in **(c)** and **(e)** show the spectral interference fringes between the LFS channels for both the classical and squeezed light pulses.

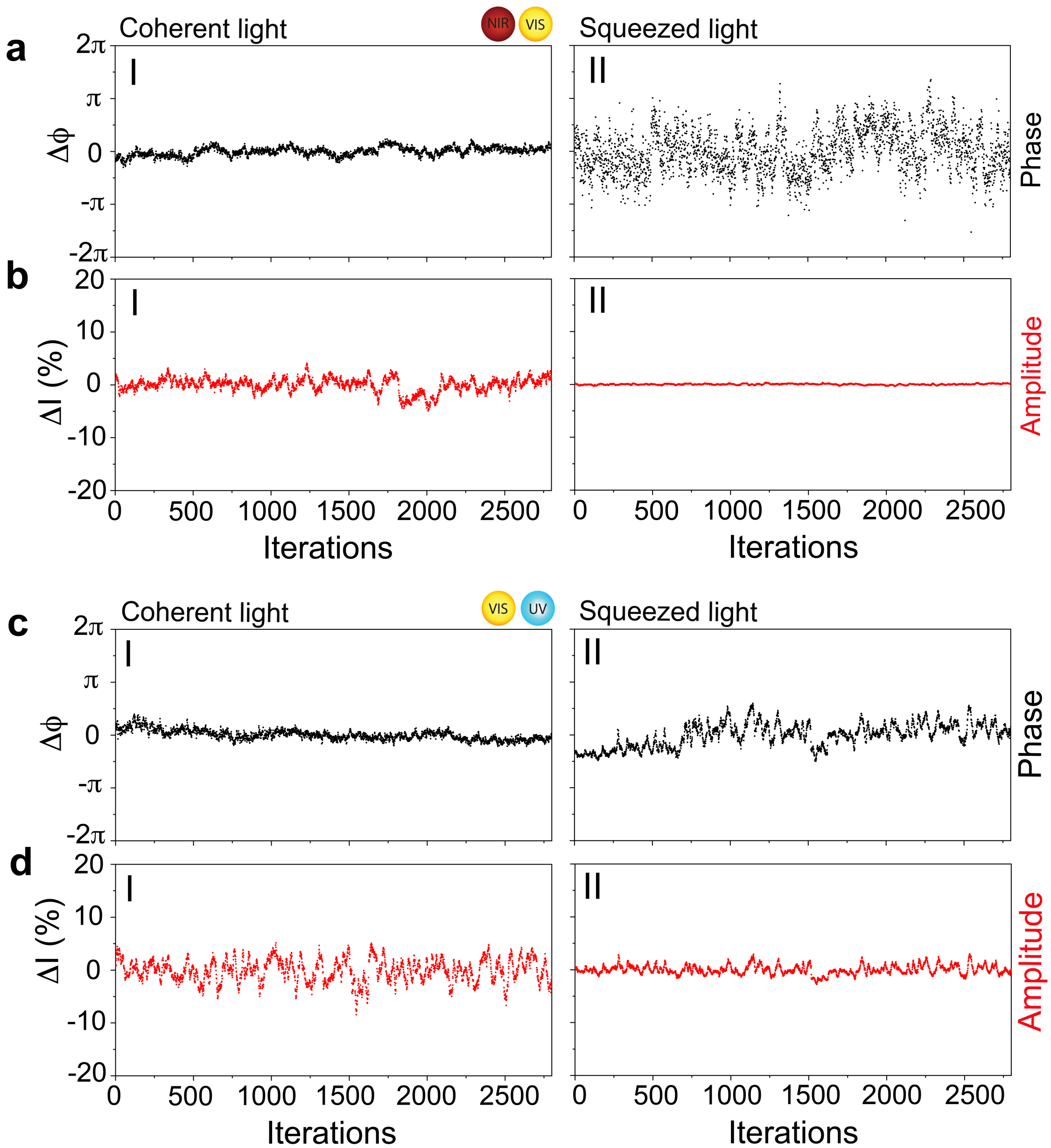

**Fig. 2. Squeezed light phase and intensity uncertainty measurements. a &b**, Measured phase (in **a**) and intensity (in **b**) uncertainties between the near-IR and visible pulses of the LFS, retrieved by averaging 2800 spectra for coherent light (**aI, bI**) and amplitude squeezed light (**aII, bII**). **c&d,** Corresponding phase and intensity uncertainty measurements for the visible and ultraviolet pulses of the LFS, presented in the same order as in panels A and B.

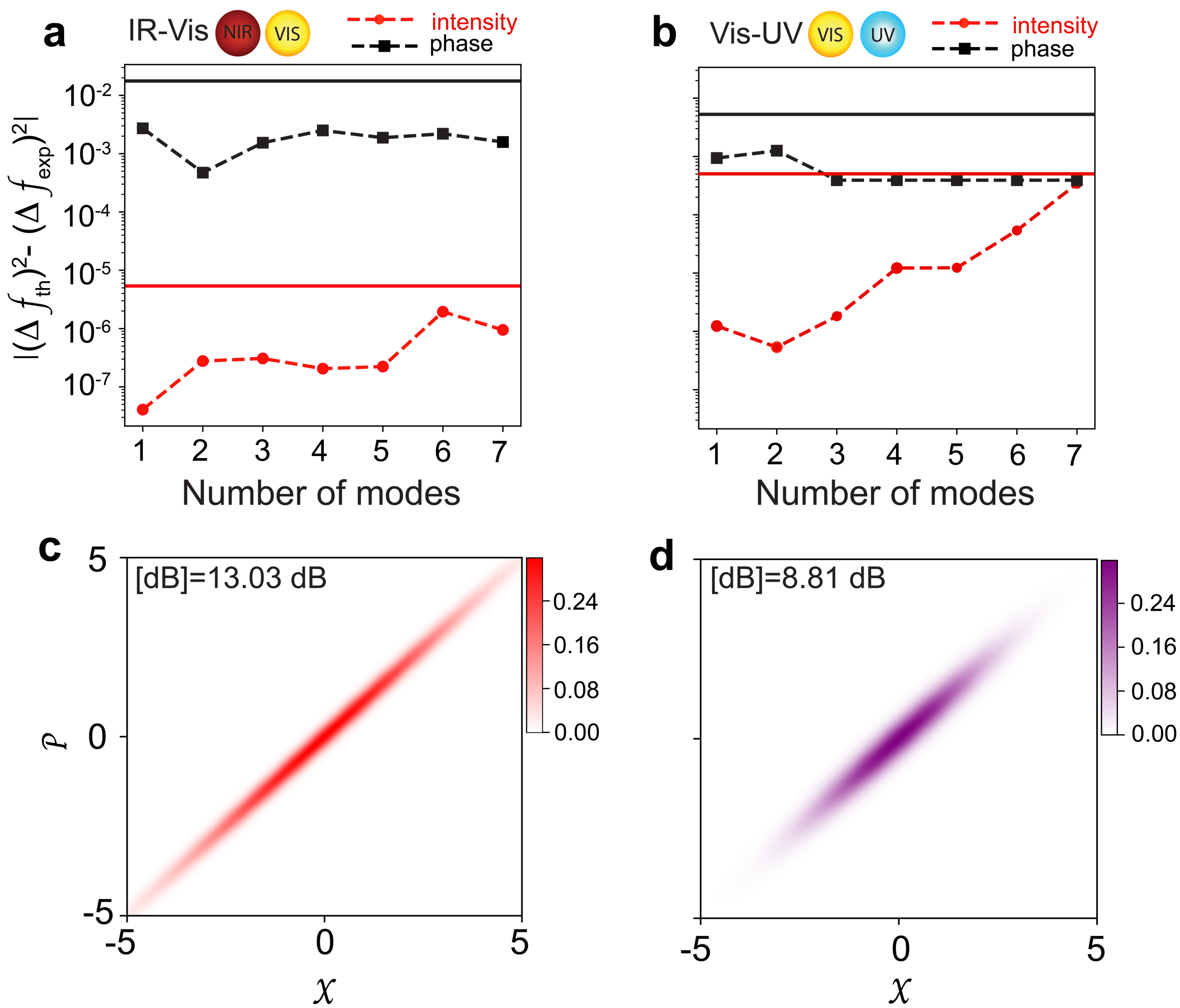

**Fig. 3. Wigner functions of the ultrafast squeezed light pulses. a &b,** The optimization results for the IR-VIS and UV-VIS datasets as a function of the number of modes. The dashed curves indicate the absolute error between the theoretical and experimental variances for intensity (in red) and phase (in black). For reference, the experimental variances are represented by solid lines in corresponding colors. **c &d,** The Wigner functions of the resulting squeezed states for the IR-VIS and UV-VIS measurements (for $N = 1$)), respectively. In these plots, x and p represent the optical quadrature, and for representational purposes, the distributions have been rotated by 45° relative to the p-axis.

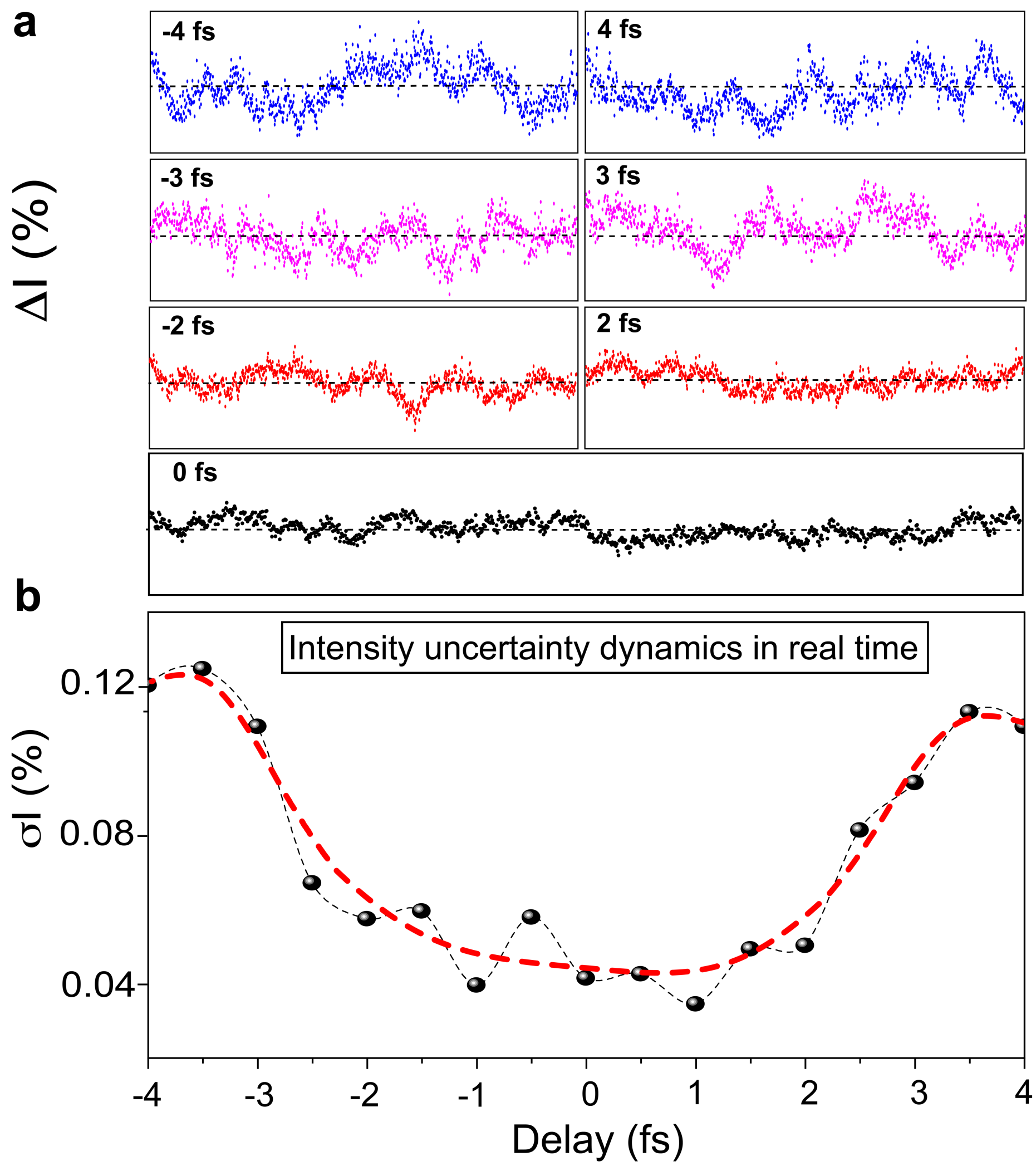

**Fig. 4. Uncertainty dynamics in real time. a,** The measured amplitude uncertainty at different delay $\tau$ times between the three input photons in the FWM nonlinear process (all are on the same scale). **b,** The amplitude uncertainty as evolves in time. The measured amplitude standard deviation "uncertainty" is shown in black dotes connected with a dashed black line (the smoothing is presented in dashed red line as a guide to the eye)

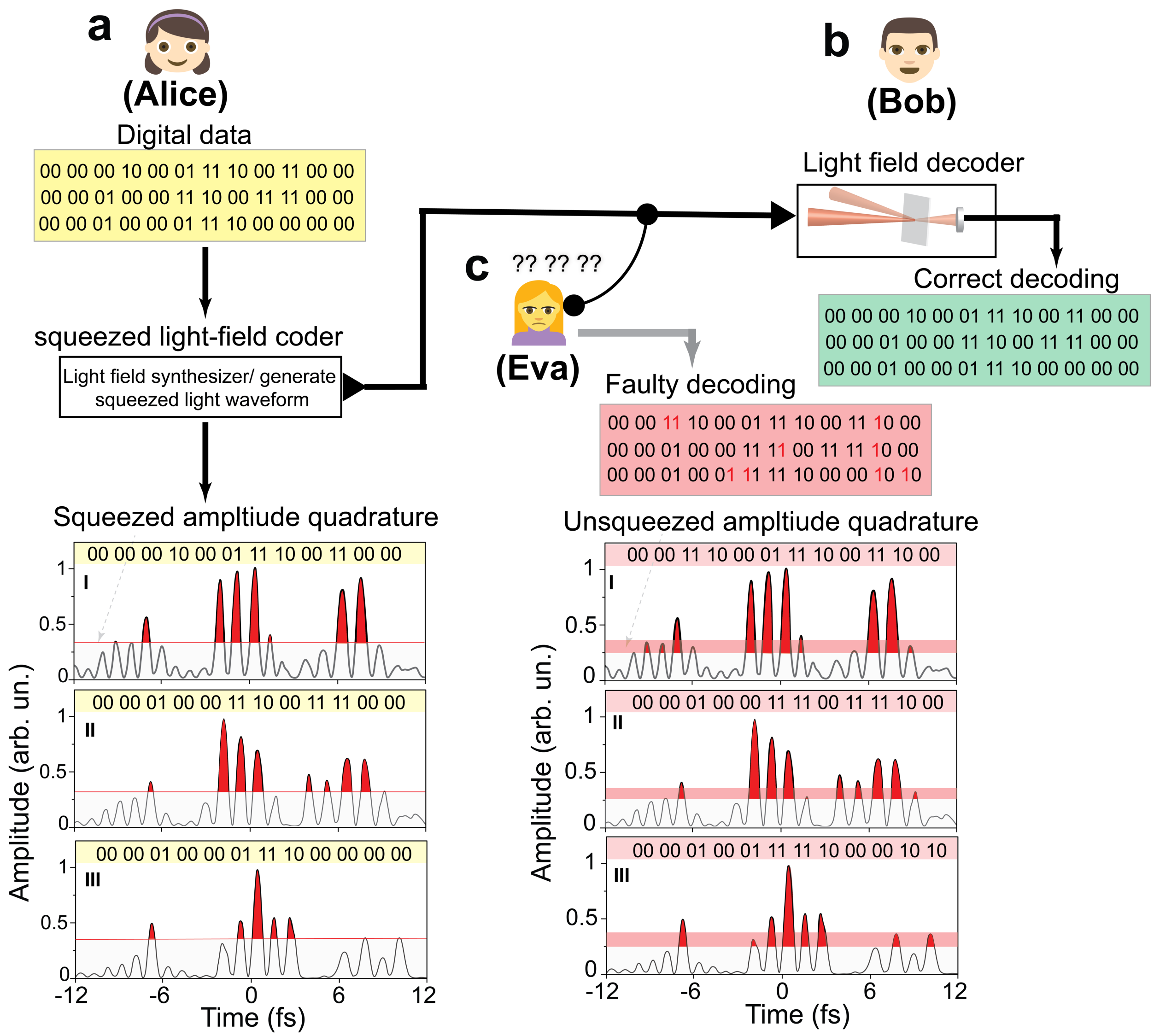

**Fig. 5. Ultrafast digital-encoded secure quantum communication. a,** Alice encodes digital data onto squeezed synthesized light waveforms (**aI-III**) using the LFS. These waveforms have predefined amplitude thresholds, indicated by the red line: signals above the threshold represent (1), and signals below represent (0). The encrypted data are carried by ultrafast squeezed light pulses, which Alice sends to Bob. She also shares the squeezing degree between channels, the predefined threshold, and the encryption key. **b,** Upon receiving the data, Bob first checks the squeezing level to confirm the security of the communication. He then decodes the data by sampling the waveform of the synthesized squeezed light pulses. **c,** Eva attempts to intercept the data using a beamsplitter. Her intervention alters the squeezing level, alerting Alice and Bob to potential tampering. Since Eva does not know the predefined threshold or the data decoding key, her attempt to decode the information results in faulty decoding, as shown in panels **cI-III**.

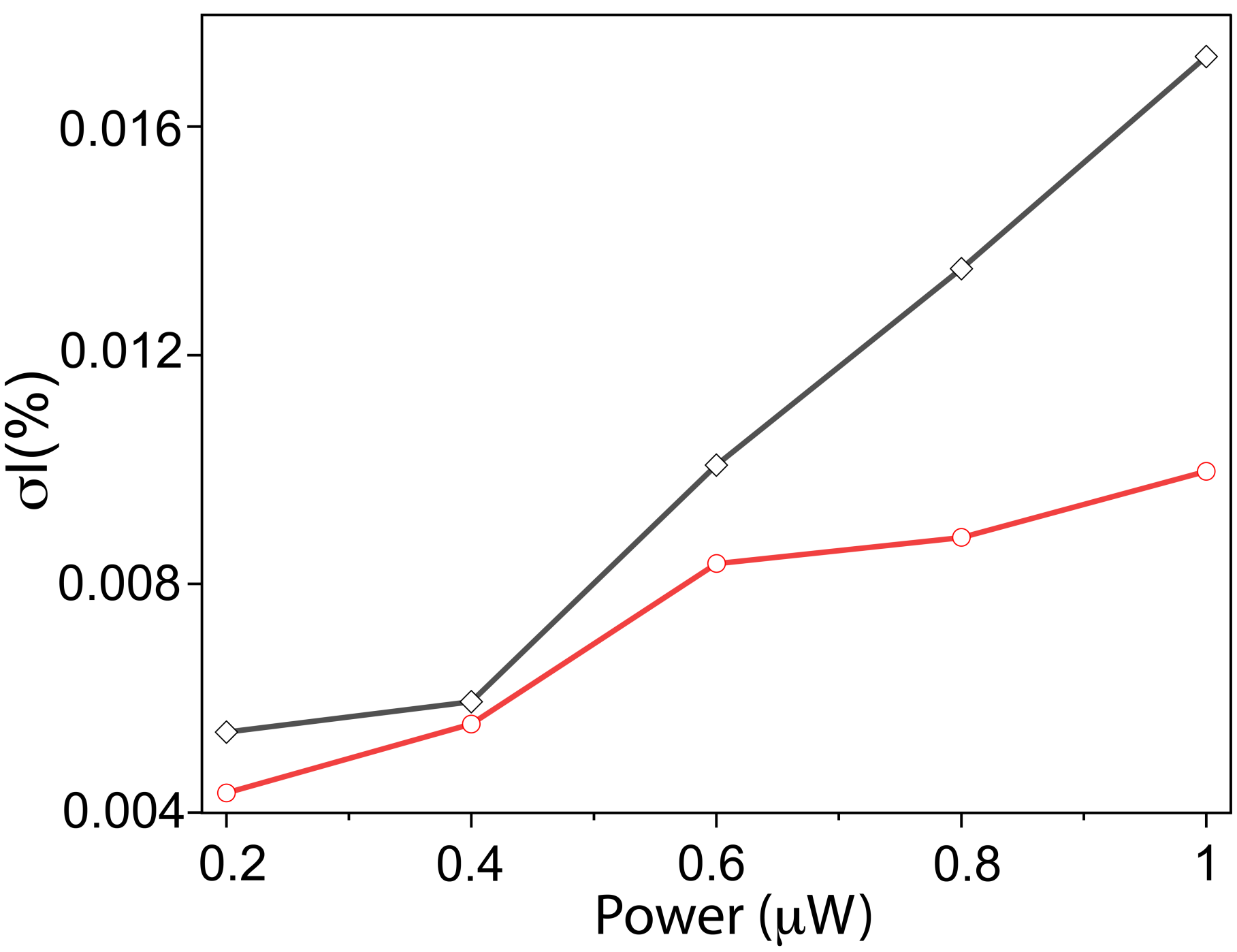

**Extended Data Fig. 1. Shot noise measurement.** The measured Δ*I* standard deviation as a function of the coherent (in back diamond point connect with black line) and amplitude squeezed (in red circles connect with red line) lights.

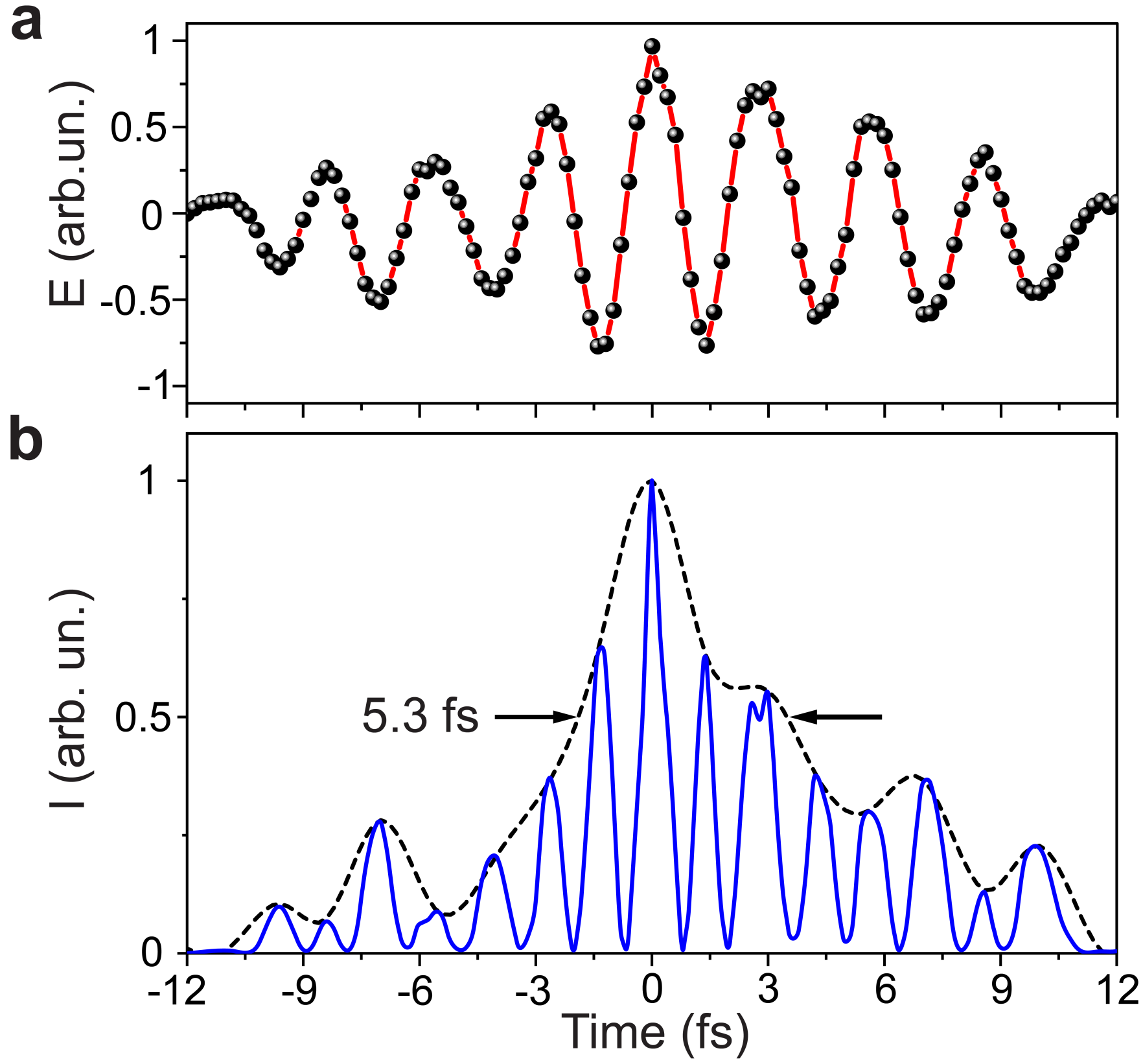

**Extended Data Fig. 2. Ultrafast squeezed light pulse. a,** Retrieved electric field, and **b,** Intensity temporal profile of the synthesized ultrafast squeezed light pulse generated by the four-wave mixing nonlinear process.

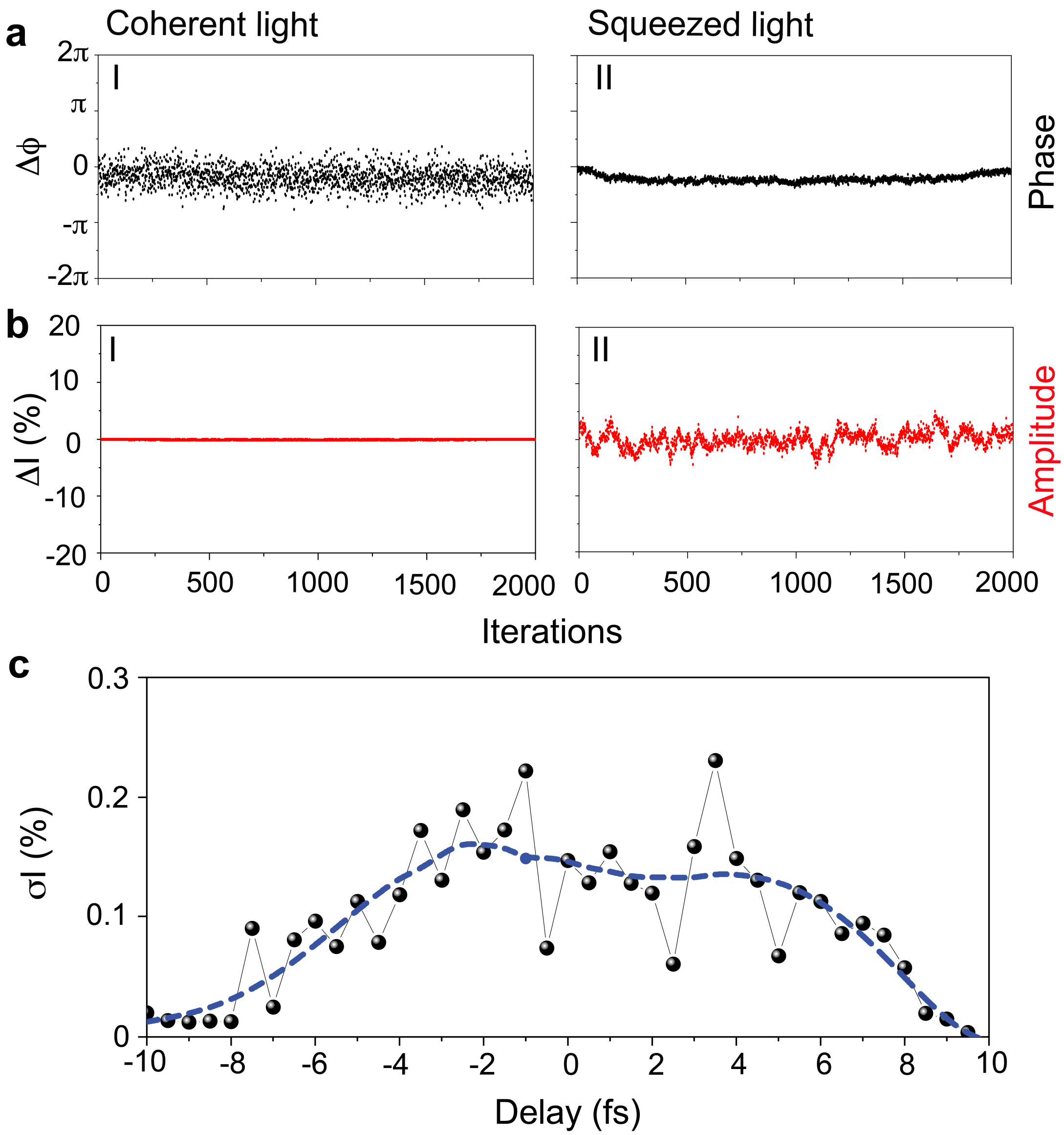

**Extended Data Fig. 3. Phase squeezed light. aI &II,** Measured phase uncertainty for coherent and phase squeezed light, between the near-IR and visible pulses of the LFS, retrieved by averaging 2000 spectra, respectively. **bI &II**, The corresponding intensity uncertainty measured for coherent, and phase squeezed light. **c,** The amplitude uncertainty of the phase squeezed light as evolves in time. The measured standard deviation (uncertainty) of the amplitude is shown in black dotes connected with a dashed black line (the smoothing is presented in dashed blue line as a guide to the eye)